\documentclass{elsart}

\usepackage{amsmath}
\usepackage{amssymb}
\usepackage{hyperref}

\DeclareMathOperator{\ov}{ov}

\begin{document}
\begin{frontmatter}
  \title{Partitions versus sets : a case of duality}

  \author{Laurent Lyaudet}
  \address{
    Universit\'e d'Orl\'eans - LIFO,\\
    Rue L\'eonard de Vinci, B.P. 6759,\\
    F-45067 Orl\'eans Cedex 2, France\\
    \texttt{\upshape Laurent.Lyaudet@gmail.com}}

  \author{Fr\'ed\'eric Mazoit\thanksref{ANR-GRAAL}}
  \address{
    Universit\'e Bordeaux - LaBRI,\\
    351, cours de la Libération,\\
    F-33405 Talence Cedex, France\\
    \texttt{\upshape Frederic.Mazoit@labri.fr}}

  \author{St\'ephan Thomass\'e\thanksref{ANR-GRAAL}}
  \address{
    Universit\'e Montpellier II - CNRS, \\
    LIRMM, 161 rue Ada, 34392 Montpellier Cedex, France\\
    \texttt{\upshape thomasse@lirmm.fr}}

  \thanks[ANR-GRAAL]{research supported by the french ANR-project
    ``Graph decompositions and algorithms (GRAAL)''}

  \begin{abstract}
    In a recent paper, Amini et al. introduced a general framework to
    prove duality theorems between tree decompositions and their dual
    combinatorial object. They unify all known ad-hoc proofs in one
    duality theorem based on submodular partition functions. This
    general theorem remains however a bit technical and relies on this
    particular submodularity property. Instead of partition functions,
    we propose here a simple combinatorial property of set of
    partitions which also gives these duality results. Our approach is
    both simpler, and a little bit more general.
  \end{abstract}
\end{frontmatter}

\section{Introduction}

In the past 30 years, several decompositions of graphs and discrete
structures such as tree-decompositions and branch-decompositions of
graphs~\cite{RoSe84a,RoSe91a}, tree-decompositions of
matroids~\cite{HlWh06a} or branch-decomposition of more general
structures~\cite{OuSe07a} have been introduced. Most of these
decompositions admit some dual combinatorial object (brambles,
tangles\dots), in the sense that a decomposition exists if and only if
the dual object does not.

In~\cite{AmMaNiTh09a}, the authors present a general framework for
proving these duality relations. Precisely, a \emph{partitioning tree}
on a finite set $E$ is a tree $T$ which leaves are identified to the
elements of $E$ in a one-to-one way. Every internal node $v$ of $T$
corresponds to the partition of $E$ which parts are the set of leaves
of the subtrees obtained by deleting $v$. Such a partition is a
\emph{node-partition}. A partitioning tree $T$ is \emph{compatible}
with a set of partitions $\mathcal{P}$ of $E$ if every node-partition
of $T$ belong to $\mathcal{P}$. For some specific sets of partitions
$\mathcal{P}$, one can get classical tree decompositions. To
illustrate our purpose, let $G=(V,E)$ be a graph (which is not too
trivial, i.e.  not a union of stars). The \emph{border} of a partition
$\mu$ of $E$ is the set of vertices incident with edges in at least
two parts of $\mu$. For every integer $k$, let $\mathcal{P}_k$ be the
set of partitions of $E$ whose border contain at most $k+1$
vertices. Now, there exists a partitioning tree compatible with
$\mathcal{P}_k$ if and only if the tree-width of $G$ is at most $k$.

The dual objects of partitioning trees are \emph{brambles}. A
\emph{$\mathcal{P}$-bramble} is a nonempty set of pairwise
intersecting subsets of $E$ which contains a part of every partition
in $\mathcal{P}$, and a $\mathcal{P}$-bramble is \emph{principal} if
it contains a singleton. A non-principal $\mathcal{P}$-bramble and a
partitioning tree compatible with $\mathcal{P}$ cannot both exist at
the same time, but there may be none of them.

In ~\cite{AmMaNiTh09a}, the authors propose a sufficient condition for
a set of partitions $\mathcal{P}$ to be such that there exists a
partitioning tree compatible with $\mathcal{P}$ if and only if no
non-principal $\mathcal{P}$-bramble does (duality property). The
condition they introduced is expressed by the mean of weight functions
on partitions. Precisely, they prove that if a partition function is
(\emph{weakly}) \emph{submodular}, the set of partitions with weight
bounded by a fixed constant enjoys the duality property. For example,
the weight function corresponding to tree-width (the size of the
border of a partition) is submodular, therefore, if the tree-width of
$G$ is more than $k$, there is no partitioning tree compatible with
$\mathcal{P}_k$, hence a bramble exists. This provides a alternative
proof of~\cite{AmMaNiTh09a}, also presented in~\cite{Di05b}. This kind
of argument provides duals for some other tree-decompositions.

Based on~\cite{AmMaNiTh09a}, Petr \v Skoda~\cite{Sk09a} studies the
complexity of computing an optimal partitioning tree and Berthom\'e
and Nisse~\cite{BeNi08a} give a unified FPT algorithm to compute a
partitioning tree but only when restricted to a subclass of submodular
partition functions.

While~\cite{AmMaNiTh09a}'s framework unifies several ad-hoc proof
techniques of duality between decompositions and their dual objects,
its core theorem mimics a proof of~\cite{RoSe91a}. The argument is
quite technical and does not give a real insight of the reason why the
duality property holds. Moreover, at least one partition function, the
function $\max_f$ which corresponds to branchwidth, is not weakly
submodular. Since this function is a limit of weakly submodular
functions, Amini et al. also manage to apply their theorem to
branchwidth but this is not truly satisfying.

The goal of this paper is twofold. First we give a simpler proof of
the duality theorem, then we slightly extend (and simplify) the
definition of \emph{weak submodularity} so that the function $\max_f$
becomes weakly submodular.

To do so, we consider \emph{partial partitioning trees}, in which the
leaves of a tree $T$ are labelled by the parts of some partition of
$E$, called the \emph{displayed partition} of $T$. When the displayed
partition consists of singletons, we have our previous definition of
partitioning trees. The set of displayed partitions of partial
partitioning trees compatible with $\mathcal{P}$ (i.e. such that every
node-partition belongs to $\mathcal{P}$) is denoted by
$\mathcal{P}^\uparrow$. Observe that in $T$, internal nodes of degree
two can be simplified, so we can assume that all internal nodes have
degree at least three.

We do not make any distinction between principal and non principal
$\mathcal{P}$-brambles. Instead we define a \emph{set of small sets}
to be a subset of $2^E$ closed under taking subset, and whose elements
are \emph{small}. We say that a set of partitions
$\mathcal{P}^\uparrow$ is \emph{dualising} if for any set of small
sets $\mathcal{S}$, there exists a \emph{big bramble} (i.e. a bramble
containing no part in $\mathcal{S}$) if and only if
$\mathcal{P}^\uparrow$ contains no \emph{small partition} (i.e a
partition whose parts all belong to $\mathcal{S}$). Thus the classical
duality results are derived when $\mathcal{S}$ consists of the empty
set and the singletons. Note that since a $\mathcal{P}$-bramble $Br$
meets all partitions in $\mathcal{P}$, if $\mathcal{P}$ contains a
small partition, $Br$ cannot contain only big parts. Hence, a class of
partitions cannot both admit a big bramble and a small partition.

In Section~\ref{sec:preliminaries}, we fix some notations and give
some basic definitions. In Section~\ref{sec:refining}, we give an
equivalent and yet easier notion than the dualising property: the
refining property. In Section~\ref{sec:pushing}, we give a sufficient
condition on $\mathcal{P}$ so that $\mathcal{P}^\uparrow$ is refining
(and thus dualising). Finally, in Section~\ref{sec:submodular}, we
extend the definition of \emph{weak submodularity} to match our
sufficient condition for duality, and we prove that the partition
function $\max_f$ is weakly submodular and thus, that branchwidth
fully belongs to the unifying framework.

\section{Brambles}
\label{sec:preliminaries}

Let $E$ be a finite set. We denote by $2^E$ the set of subsets of
$E$. A \emph{partition} of $E$ is a set of pairwise disjoint subsets
of $E$ which cover $E$ and whose parts are non empty. The sets
$\mathcal{P}$ and $\mathcal{Q}$ denote sets of partitions of
$E$. Greek letters $\alpha$, $\beta$,\dots denote sets of nonempty
subsets of $E$, while capital letters $A$, $B$,\dots denote nonempty
subsets of $E$. We write $X^c$ for the complement $E\setminus X$ of
$X$. We denote a finite union
$\alpha_1\cup\alpha_2\cup\dots\cup\alpha_p$ by
$(\alpha_1|\alpha_2|\dots|\alpha_p)$ and also shorten
$(\{A\}|\alpha|\{B\})$ into $(A|\alpha|B)$. The \emph{size} of a
subset $\alpha$ of $2^E$ is just the number of sets in $\alpha$.  For
any $F$, $\alpha\setminus F$ denotes the set $\{A\setminus F\;;\; A\in
\alpha\}$, where empty sets have been removed. The \emph{overlap} of
$\alpha$ is the set $\ov(\alpha)$ of the elements that belong to at
least two parts of $\alpha$.

Let $T$ and $T'$ be two partial partitioning trees respectively
displaying $(\alpha|A)$ and $(A^c|\beta)$ with $u$ a leaf of $T$
labelled $A$ and $u'$ a leaf of $T'$ labelled $A^c$. Take the disjoint
union of $T$ and $T'$. Link the respective neighbours of $u$ and $u'$
and remove $u$ and $u'$. What we get is a new partitioning tree which
displays $(\alpha|\beta)$.  We say that $(\alpha|\beta)$ is the
\emph{merged} partition of $(\alpha|A)$ and $(A^c|\beta)$. It is easy
to check that the set $\mathcal{P}^\uparrow$ of all displayed
partitions of partial partitioning trees is exactly the least superset
of $\mathcal{P}$ which is closed under merging of partitions.

\begin{lem}\label{lem:decompose}
  For any $(\alpha|A)\in\mathcal{P}^\uparrow\setminus \mathcal{P}$,
  there exists $(\gamma|C)\in\mathcal{P}$ and
  $(C^c|\mu|A)\in\mathcal{P}^\uparrow$ such that
  $(\alpha|A)=(\gamma|\mu|A)$, where $(\gamma|C)$ has at least three
  parts.
\end{lem}
\begin{pf}
  Let $T$ be some partial partitioning tree which displays
  $(\alpha|A)$.  Since $(\alpha|A)$ does not belong to $\mathcal{P}$,
  $T$ has at least two internal nodes.

  The partition $(\gamma|C)$ can be any node-partition of an internal
  node of $T$ which is adjacent to only one internal node and not
  adjacent to the leaf $A$.
  \hfill\strut\qed
\end{pf}

We say that such a partition $(\gamma|C)$ \emph{decomposes}
$(\alpha|A)$. To extend this notion to $\mathcal{P}^\uparrow$, we also
say that $(\alpha|A)$ \emph{decomposes} $(\alpha|A)$, when
$(\alpha|A)\in\mathcal{P}$.

Starting with some subset $\beta$ of $2^E$, one can perform two
operations:
\begin{itemize}
\item (Deletion) Suppress an element in some set of
  $\beta$. Precisely, if $\beta=(B|\gamma)$ and $b\in B$, the result
  of the deletion operation is $(B\setminus \{b\}|\gamma)$.
\item (Partition) Partition some set of $\beta$. Precisely, if
  $\beta=(B|\gamma)$ and $\delta$ is a partition of $B$, the result of
  the partition operation is $(\delta|\gamma)$.
\end{itemize}

We say that $\alpha$ is \emph{finer} than $\beta$ if it can be
obtained from $\beta$ by a sequence of deletions and
partitions. Observe that in some cases, the deletion operation can
result in an empty set. In these cases, since we do not allow the
empty set in our families of sets, we simply delete the set. When we
write that $(\alpha_1|\dots|\alpha_p)$ is finer than
$(\beta_1|\dots|\beta_q)$, with $p\leq q$, we usually mean that each
$\alpha_i$ is finer than $\beta_i$. Note that if $\alpha$ is finer
than $\beta$, then $\ov(\alpha)$ is included in $\ov(\beta)$.

A \emph{$\mathcal{P}$-bramble}, or just \emph{bramble} when no
confusion can occur, is a set $Br$ of subsets of $E$ such that
\begin{itemize}
\item $Br$ contains a part of every $\mu\in\mathcal{P}$ ($Br$
  \emph{meets} every $\mu\in\mathcal{P}$);
\item the elements of $Br$ are pairwise intersecting.
\end{itemize}
If $Br$ is a $\mathcal{P}$-bramble, we say that $\mathcal{P}$
\emph{admits} the bramble $Br$.

A set $\mathcal{S}$ of \emph{small} sets is just a subset of $2^E$
which is closed under taking subsets. A set which does not belong to
$\mathcal{S}$ is \emph{big}. By extension, a \emph{big bramble} is a
bramble consisting exclusively of big sets, while a \emph{small
  partition} only contains small parts.

If we consider directly $\mathcal{P}^\uparrow$, we have a dummy
duality theorem which states that: Either there is a small partition
in $\mathcal{P}^\uparrow$, or there is a set containing a big part of
every $\mu\in\mathcal{P}^\uparrow$.  Thus the pairwise intersection
condition is not required.  However this condition is necessary to
restrict the obstruction to $\mathcal{P}$.

\begin{lem}\label{rem:1}
  A set $Br$ is a $\mathcal{P}$-bramble if and only if it is a
  $\mathcal{P}^\uparrow$-bramble.
\end{lem}
\begin{pf}
  Let $Br$ be a set of subsets of $E$. Since
  $\mathcal{P}\subseteq\mathcal{P}^\uparrow$, if $Br$ is a
  $\mathcal{P}^\uparrow$ bramble, then $Br$ is a $\mathcal{P}$ bramble
  too. Now suppose that $Br$ is not a $\mathcal{P}^\uparrow$-bramble.
  If $Br$ contains disjoint elements, it cannot be a
  $\mathcal{P}$-bramble so let us suppose that $Br$ contains no part
  of some partition $\mu\in\mathcal{P}^\uparrow$. Take $\mu$ with
  minimum number of parts. If $\mu\in\mathcal{P}$, then $Br$ is not a
  $\mathcal{P}$-bramble, otherwise $\mu=(\alpha|\beta)$ for some
  $(\alpha|A)$, $(A^c|\beta)\in\mathcal{P}^\uparrow$ has less parts
  than $\mu$. Since $\mu$ is minimal, $Br$ contains a part of both
  $(\alpha|A)$ and $(A^c|\beta)$ and no part of $(\alpha|\beta)$. It
  contains both $A$ and $A^c$ which are disjoint, and thus $Br$ is not
  a bramble.
  \hfill\strut\qed
\end{pf}

\section{Dualising and refining sets of partitions}
\label{sec:refining}

We will only apply the theorems of this section to sets of partitions
of the form $\mathcal{P}^\uparrow$, but since these results are valid
in the general case, we express them for any set $\mathcal{Q}$ of
partitions of $E$.

A set of partitions $\mathcal{Q}$ is \emph{dualising} if for any set
of small sets $\mathcal{S}$, either there exists a big
$\mathcal{Q}$-bramble, or $\mathcal{Q}$ contains a small partition.

A set of partitions $\mathcal{Q}$ is \emph{refining} if for any
$(\alpha|A)$, $(B|\beta)\in\mathcal{Q}$ with $A$ disjoint from $B$,
there exists a partition in $\mathcal{Q}$ which is finer than the
covering $(\alpha|\beta)$.

\begin{thm}\label{th:duality1}
  If $\mathcal{Q}$ is refining, then $\mathcal{Q}$ is dualising.
\end{thm}
\begin{pf}
  Suppose that $\mathcal{Q}$ is refining and that $\mathcal{Q}$
  contains no small partition for some set of small sets. There exists
  a set that contains a big part from every partition in
  $\mathcal{Q}$, and which is closed under taking superset (just
  consider the set of all big sets). We claim that such a set $Br$,
  chosen inclusion-wise minimal, is a big bramble.

  If not, there exists two disjoint sets $A$ and $B$ in $Br$. Choose
  them inclusion-wise minimal. Since $Br\setminus\{A\}$ is upward
  closed and $Br$ is minimal, there exists $(\alpha|A)\in\mathcal{Q}$
  which contains no part of $Br\setminus\{A\}$. Similarly, there
  exists $(B|\beta)\in\mathcal{Q}$ which contains no part of
  $Br\setminus\{B\}$. Hence $Br$ does not meet $(\alpha|\beta)$, but
  since $\mathcal{Q}$ is refining, it contains a partition which is
  finer than $(\alpha|\beta)$ and which is not met by $Br$, a
  contradiction.  \hfill\strut\qed
\end{pf}

Conversely,

\begin{thm}\label{th:duality2}
  If $\mathcal{Q}$ is dualising, then $\mathcal{Q}$ is refining.
\end{thm}
\begin{pf}
  Assume for contradiction that $\mathcal{Q}$ is not refining. Let
  $(\alpha|A)$ and $(B|\beta)\in\mathcal{Q}$ with $A$ and $B$ disjoint
  and such that $\mathcal{Q}$ contains no partition which is finer
  than $(\alpha|\beta)$.  Choose, as small sets, all the sets included
  in some part of $(\alpha|\beta)$.
  \begin{itemize}
  \item Since $\mathcal{Q}$ contains no partition which is finer than
    $(\alpha|\beta)$, there is no small partition.
  \item Since a bramble $Br$ cannot both contain $A$ and $B$, to meet
    both $(\alpha|A)$ and $(B|\beta)$, it must contain a small
    set. Thus $Br$ cannot be a big bramble.
  \end{itemize}
  This proves that $\mathcal{Q}$ is not dualising. \hfill\strut\qed
\end{pf}

We would like to emphasise that in the following, we only use
Theorem~\ref{th:duality1}.

\section{Pushing sets of partitions}
\label{sec:pushing}

We now introduce a property on $\mathcal{P}$ which implies that
$\mathcal{P}^\uparrow$ is refining and thus, by
Theorem~\ref{th:duality1}, that $\mathcal{P}^\uparrow$ is dualising.

A set of partitions $\mathcal{P}$ is \emph{pushing} if for every pair
of partitions $(\alpha|A)$ and $(B|\beta)$ in $\mathcal{P}$ with
$A^c\cap B^c\neq\emptyset$, there exists a nonempty $F\subseteq
A^c\cap B^c$ such that $(\alpha\setminus F|A\cup F)\in\mathcal{P}$ or
$(B\cup F|\beta\setminus F)\in\mathcal{P}$.

To prove that if $\mathcal{P}$ is pushing, then $\mathcal{P}^\uparrow$
is refining, we have to strengthen the refining property as
follows. If a partition $\alpha$ is only obtained from $\beta$ by
deletions, we say that $\alpha$ is \emph{strongly finer} than $\beta$,
and a set $\mathcal{Q}$ of partition of $E$ is \emph{strongly
  refining} if for any $(\alpha|A)$, $(B|\beta)\in\mathcal{Q}$ with
$A$ disjoint from $B$, there exists a partition in $\mathcal{Q}$
strongly finer than the covering $(\alpha|\beta)$. Clearly if
$\mathcal{Q}$ is strongly refining, then it is refining, the following
theorem thus implies that if $\mathcal{P}$ is pushing, then
$\mathcal{P}^\uparrow$ is refining.

\begin{thm}\label{th:push->strong-refining}
  If $\mathcal{P}$ is pushing, then $\mathcal{P}^\uparrow$ is strongly
  refining.
\end{thm}
\begin{pf}
  Suppose for a contradiction that $\mathcal{P}$ is pushing, that
  $(\alpha|A)$, $(B|\beta)$ both belong to $\mathcal{P}^\uparrow$ with
  $A$ disjoint from $B$, and yet $\mathcal{P}^\uparrow$ contains no
  partition strongly finer than $(\alpha|\beta)$. Choose
  $(\alpha|\beta)$ with minimum number of parts, and then with minimum
  overlap among counter-examples with minimal size. Let $O=A^c\cap
  B^c$ be the overlap of $(\alpha|\beta)$. Observe that since
  $(\alpha|\beta)$ is not a partition of $E$, $O$ is nonempty.

  We claim that there exist no $(\gamma|C)$,
  $(D|\delta)\in\mathcal{P}^\uparrow$ with $C$ disjoint from $D$, such
  that $(\gamma|\delta)$ is strongly finer than $(\alpha|\beta)$ and
  has an overlap which is a strict subset of $O$. If not, our choice
  of $(\alpha|A)$, $(B|\beta)$ implies that $\mathcal{P}^\uparrow$
  contains a partition $\lambda$ which is strongly finer than
  $(\gamma|\delta)$ and thus $\lambda$ is strongly finer than
  $(\alpha|\beta)$, a contradiction.

  By Lemma~\ref{lem:decompose}, let $(\gamma|C)$ and $(D|\delta)$ be
  respectively decomposing $(\alpha|A)$ and $(B|\beta)$. Since
  $A\subseteq C$ and $B\subseteq D$, we have $C^c\cap D^c\subseteq
  O$. If $C^c\cap D^c$ is nonempty, since $\mathcal{P}$ is pushing,
  there exists a nonempty subset $F$ of $O$ such that, say,
  $(\gamma\setminus F,C\cup F)\in\mathcal{P}$. If $C^c$ and $D^c$ are
  disjoint, they cannot both contain $O$. There thus exists a non
  empty $F\subseteq O$ which is disjoint from, say, $C^c$, and
  therefore $(\gamma|C)=(\gamma\setminus F, C\cup F)$. In both cases,
  $(\gamma\setminus F, C\cup F)\in\mathcal{P}$.
  \begin{itemize}
  \item If $(\gamma|C)=(\alpha|A)$, then $(\gamma\setminus F|\beta)$
    is strongly finer than $(\alpha|\beta)$ and its overlap is
    $O\setminus F$, which is strictly included in $O$, a
    contradiction.
  \item If $(\gamma|C)\neq(\alpha|A)$, we consider
    $(C^c|\mu|A)\in\mathcal{P}^\uparrow$ such that
    $(\gamma|\mu|A)=(\alpha|A)$. Since $(C^c|\mu|A)$ has less parts
    than $(\alpha|A)$, there exists
    $(C'|\mu'|\beta')\in\mathcal{P}^\uparrow$ which is strongly finer
    than $(C^c|\mu|\beta)$. We assume that $C'\subseteq C^c$ is
    nonempty, since $(\mu'|\beta')\in\mathcal{P}^\uparrow$ would be
    strongly finer than $(\alpha|\beta)$. If $O\not\subseteq C'^c$,
    then $(\gamma|\mu'|\beta')$ is strongly finer than
    $(\alpha|\beta)$, with an overlap strictly included in $O$, a
    contradiction. If $O\subseteq C'^c$, then $C'$ and $C\cup F$ are
    disjoint. But then $(\gamma \setminus F|\mu'|\beta')$ is strongly
    finer than $(\alpha|\beta)$, and its overlap (which is a subset of
    $O\setminus F$) is a strict subset of $O$, a contradiction.
    \hfill\strut\qed
  \end{itemize}
\end{pf}

Observe that $\mathcal{P}$ being pushing implies
$\mathcal{P}^\uparrow$ begin refining, but we could not avoid the
strong version of refinement in our proof. For instance, the relaxed
statement of theorem~\ref{th:push->strong-refining}, with refining
only, makes the first claim of the proof to fail. We could imagine
that there exists $(\gamma|C)$ and $(D|\delta)\in\mathcal{P}^\uparrow$
with $C$ and $D$ disjoint, $(\gamma|\delta)$ finer than
$(\alpha|\beta)$ with a smaller overlap but with more parts than
$(\alpha|\beta)$.

\section{Submodular partition functions}
\label{sec:submodular}

A \emph{partition function} is a function from the set of partitions
of $E$ into $\mathbb{R}\cup\{+\infty\}$. In~\cite{AmMaNiTh09a}, the
authors define \emph{submodular} partition functions $\Psi$ such that
for every partitions $(\alpha|A)$ and $(B|\beta)$, we have:
\begin{equation*}
  \Psi(\alpha|A)+\Psi(B|\beta)
  \geq\Psi(\alpha \setminus B^c|A\cup B^c)
  +\Psi(\beta \setminus A^c|B\cup A^c).
\end{equation*}

It is routine to observe that if $\Psi $ is partition submodular, then
for every $k$, the set $\mathcal{P}_k$ of partitions with $\Psi$ value
at most $k$ is pushing, just consider for this $F=A^c\cap B^c$ in the
definition of the pushing property. Hence $\mathcal{P}_k^\uparrow$ is
dualising as soon as $\Psi$ is submodular. From this follows the
duality theorems for tree-width of matroids and graphs, as explicited
in~\cite{AmMaNiTh09a}.

However, in order to also obtain duality for branchwidth, the authors
introduce \emph{weakly submodular} partition functions as partition
functions such that for every partitions $(\alpha|A)$ and $(B|\beta)$,
at least one of the following holds:
\begin{itemize}
\item there exists $A\subset F\subseteq (B\setminus A)^c$ with
  $\Psi((\alpha|A)) > \Psi((\alpha\setminus F|A\cup F))$;
\item $\Psi((\beta|B)) \geq \Psi((\beta\setminus A^c|B\cup A^c))$.
\end{itemize}

Since $(\beta|B)$ and $(\beta\setminus A^c|B\cup A^c)$ are equal when
$A^c\cap B^c=\emptyset$, this definition is only really interesting
when $A^c\cap B^c\neq\emptyset$.

We introduce now a more convenient property, still called \emph{weak
  submodularity}, in which partition functions satisfy that for every
$(\alpha|A)$ and $(B|\beta)$ with $A^c\cap B^c\neq\emptyset$, there
exists a nonempty $F\subseteq A^c\cap B^c$ such that at least one of
the following holds:
\begin{itemize}
\item $\Psi((\alpha|A))\geq \Psi((\alpha\setminus F|A\cup F))$;
\item $\Psi((\beta|B))\geq \Psi((\beta\setminus F|B\cup F))$.
\end{itemize}

This definition indeed generalises the previous one.
\begin{itemize}
\item Suppose that there exists $A\subset F\subseteq (B\setminus A)^c$
  with $\Psi((\alpha|A)) > \Psi((\alpha\setminus F|A\cup F))$. Set
  $F':=F\cap (A^c\cap B^c)$. Since $F=F'\cup A$, $(\alpha\setminus
  F|A\cup F)=(\alpha\setminus F'|A\cup F')$. Thus $\Psi((\alpha|A)) >
  \Psi((\alpha\setminus F'|A\cup F'))$ and $F'$ is certainly nonempty.
\item Suppose that $\Psi((\beta|B))\geq \Psi((\beta\setminus A^c|B\cup
  A^c))$. Set $F:=A^c\cap B^c$. Since $(\beta\setminus A^c|B\cup
  A^c)=(\beta\setminus F|B\cup F)$, $\Psi((\beta|B))\geq
  \Psi((\beta\setminus F|B\cup F))$ and $F$ is nonempty.
\end{itemize}

\begin{claim}
  A set of partition $\mathcal{P}$ is pushing if and only if
  $\mathcal{P}=\{\mu\;;\;\Psi(\mu)\leq k\}$ for some weakly submodular
  partition function $\Psi$ and $k\in\mathbb{R}\cup\{+\infty\}$.
\end{claim}
Obviously given a weakly submodular partition function $\Psi$, the
class of partitions $\mathcal{P}_k = \{\alpha \;;\; \Psi(\alpha) \leq
k\}$, for some $k \in \mathbb{R}$, is pushing. Conversely if
$\mathcal{P}$ is pushing, then defining $\Psi$ as $\Psi(\alpha)=0$ if
$\alpha \in \mathcal{P}$ and $\Psi(\alpha)=1$ otherwise, we obtain a
weakly submodular partition function.

A \emph{connectivity function} is a function $f:2^E\mapsto
\mathbb{R}\cup\{+\infty\}$ which is \emph{symmetric} (i.e. for any
$A\subseteq E$, $f(A)=f(A^c)$) and \emph{submodular} (i.e. for any
$A$, $B\subseteq E$, $f(A)+f(B)\geq f(A\cup B)+f(A\cap B)$). For any
connectivity function $f$, we define the partition function $\max_f$
by $\max_f(\alpha)=\max\{f(A) \;;\; A\in\alpha\}$ ($\alpha$ a
partition of $E$). The weak submodularity of the $\max_f$ function
gives the duality theorems concerning branchwidth and rankwidth.

\begin{lem}\label{lem:maxf_weakly}
  The function $\max_f$ is a weakly submodular partition function.
\end{lem}
\begin{pf}
  Let $(\alpha|A)$ and $(B|\beta)$ be two partitions of $E$ such that
  $A^c\cap B^c$ is nonempty. Let $F$ with $A\setminus B\subseteq
  F\subseteq (B\setminus A)^c$ be such that $f(F)$ is minimum. We
  claim that
  $\mathrm{max}_f((\alpha|A))\geq\mathrm{max}_f((\alpha\setminus
  F|A\cup F))$.

  Indeed, we have $f(F\cap A)\geq f(F)$ by definition of $F$, and by
  submodularity, since $f(F)+f(A)\geq f(A\cap F)+f(A\cup F)$, we have
  $f(A)\geq f(A\cup F)$. For every $X$ in $\alpha$, we have by
  submodularity of $f$:
  \begin{equation}\label{truc1}
    f(X)+f(F^c) \geq f(X\cap F^c) + f(X\cup F^c)
  \end{equation}

  Since $f(F)$ is minimum, $f(F)\leq f(F\setminus X)$, and thus $f$
  being symmetric:
  \begin{equation}\label{truc2}
    f(X\cup F^c)\geq f(F^c)
  \end{equation}

  Adding \eqref{truc1} and \eqref{truc2}, we obtain $f(X)\geq f(X\cap
  F^c)$. Thus $\max_f((\alpha|A)) \geq \max_f((\alpha\setminus F,A\cup
  F))$, as claimed.

  Similarly, $\max_f((B|\beta)) \geq \max_f((B\cup F^c|\beta\setminus
  F^c))$. Now at least one of $F_A:=F\cap (A^c\cap B^c)$ and
  $F_B:=F^c\cap (A^c\cap B^c)$, say $F_A$, is nonempty. Since
  $(\alpha\setminus F|A\cup F)=(\alpha\setminus F_A|A\cup F_A)$, there
  exists a nonempty $F_A\subseteq A^c\cap B^c$ with
  $\max_f((\alpha|A)) \geq \max_f((\alpha\setminus F_A,A\cup F_A))$
  which proves that $\max_f$ is weakly submodular.  \hfill\strut\qed
\end{pf}

Together with Theorems~\ref{th:duality1}
and~\ref{th:push->strong-refining}, Lemma~\ref{lem:maxf_weakly} gives
a new proof of the branchwidth and rankwidth duality theorems.

\section{Conclusion}
\label{sec:conclusion}

In the present paper, we solve some shortcomings of~\cite{AmMaNiTh09a}
by changing a bit the original framework and, mainly, by
exhibiting a specific property of sets of partitions instead
of defining these sets via the use of partition function. Here 
are some points in which our approach differs significantly:
\begin{itemize}
\item In~\cite{AmMaNiTh09a}, the ``interesting'' brambles are the
  non-principal ones. These brambles do not contain elements that
  appear as leaves of partitioning trees, i.e. singletons. The duality
  property thus relates partitioning trees and non-principal brambles.

  In the present paper we relax the condition on the leaves of a
  partitioning tree by only requiring that these are small sets. In
  this setting, the duality property relates partial partitioning
  trees displaying a small partition to big brambles.

\item By introducing the refinement property, we give an equivalent
  version of the dualising property. This simplifies the
  technicalities of the previous proofs, as well as it highlights the
  fact that this dualising/refinement property is a natural definition
  in the study of sets of partitions.

\item Finally, the previous definition of weak submodularity being not
  entirely satisfactory (lack of symmetry, problem with branchwidth)
  we propose a new definition which simplifies and unifies the previous one.
\end{itemize}

\end{document}